\documentstyle[12pt,epsfig]{article}
\newlength{\dinwidth}
\newlength{\dinmargin}
\newcommand{\ba}{\begin{array}}
\newcommand{\ea}{\end{array}}
\newcommand{\be}{\begin{equation}}
\newcommand{\ee}{\end{equation}}
\newcommand{\bear}{\begin{eqnarray}}
\newcommand{\eear}{\end{eqnarray}}
\newcommand{\gsim}{\mathrel{\mathop{\kern 0pt \rlap
  {\raise.2ex\hbox{$>$}}} \lower.9ex\hbox{\kern-.190em $\sim$}}}

\def\ben{\begin{equation}}
\def\een{\end{equation}}
\def\bea{\begin{eqnarray}}
\def\eea{\end{eqnarray}}
\def\nn{\nonumber}
\def\half{{1\over 2}}

\begin{document}
\thispagestyle{empty}
\addtocounter{page}{-1}
\vskip-0.35cm
\begin{flushright}
UK-04-20 \\
\end{flushright}
\vspace*{0.2cm}
\centerline{\Large \bf Spacelike boundaries from the $c=1$ Matrix Model}
\vspace*{1.0cm} 
\centerline{{\bf Sumit R. Das${}^{a}$} and  {{\bf Joanna L. Karczmarek${}^b$}}}
\vspace*{0.7cm}
\centerline{ ${}^a$\it Department of Physics and Astronomy,}
\vspace*{0.2cm}
\centerline{\it University of Kentucky, Lexington, KY 40506 \rm USA}
\vspace*{0.35cm}
\centerline{${}^b$\it Jefferson Physical Laboratory}
\vspace*{0.2cm}
\centerline{\it Harvard University, Cambridge, MA 02138 \rm USA}
\vspace*{1cm}
\centerline{\tt das@gravity.pa.uky.edu
\hskip0.7cm karczmarek@physics.harvard.edu}

\vspace*{0.8cm}

\centerline{\bf Abstract}
\vspace*{0.3cm}
We find classical solutions of two dimensional noncritical string
theory which give rise to geometries with {\em spacelike} boundaries, 
similar to spacetimes with cosmological event horizons.
In the $c=1$ matrix model, these solutions have a representation as 
simple time dependent configurations.  We obtain the causal structure of
the resulting spacetimes.  Using the macroscopic loop transform,
we probe the form of the tachyon condensate in the asymptotic regions.

\vspace*{0.5cm}

\baselineskip=18pt
\newpage

\section{Introduction}

The $c=1$ matrix model \cite{ceqone}, interpreted as a theory of D0 branes 
\cite{McGreevy:2003kb,Klebanov:2003km}, is the holographic description of
two dimensional string theory. It provides an example of open-closed
duality with an explicit dictionary, since the density of matrix 
eigenvalues is directly related to the string field \cite{Das:1990ka}. 
The availability of such a model has inspired recent work  
aimed at understanding the physics of time
dependent backgrounds. While classical time dependent solutions
of the matrix model have been known for quite some time 
\cite{Minic:1991rk,Moore:1992gb,Alexandrov:2002fh}, 
only recently has it  been suggested
\cite{Karczmarek:2003pv} that these backgrounds should be regarded as
``matrix cosmologies'' and fruitfully utilized to understand
conceptual issues related to quantum cosmology such as particle
production \cite{Karczmarek:2004ph,Das:2004hw,Mukhopadhyay:2004ff}
and thermality \cite{Karczmarek:2004yc}.

The most closely studied class of time dependent solutions
\cite{Karczmarek:2004ph}-\cite{Karczmarek:2004yc}
is one in which
the Liouville wall accelerates toward the weakly coupled
region, and approaches ${\cal I}^+$ on an asymptotically null trajectory.
In the fermi sea picture this was described by the edge of the fermi
sea of eigenvalues moving away to infinity.

In this paper, we discuss a different class of solutions where
the edge of the fermi sea  {\em disappears} after some time,
and the left and the right seas merge. 
Remarkably, we will find that in spacetime this
is accompanied by the appearance of a future ${\cal I}^+$ 
which is {\em spacelike} rather than {\em null}. A time
reversed version similarly leads to a spacelike ${\cal I}^-$ while a
time-symmetric version renders both ${\cal I}^+$ and ${\cal I}^-$
spacelike.  

While the causal structure of the spacetime
is determined, computing the conformal factor
of the metric is a much more delicate task which we do 
not attempt.  The metric, as will be discussed, cannot
be computed from classical information contained
in the effective action for the collective field.
A comparison of the computation of a {\em quantum effect}
from the effective action 
for string theory with that from the matrix model, 
e.g. the computation of the outgoing stress-energy tensor of 
particles produced in the nontrivial background, might shed
some light on this question \cite{Karczmarek:2004bw}, 
but is well beyond the scope of this paper.

It is well known that the bulk spacetime which naturally follows from
the matrix model is related to the spacetime of perturbative string
theory by a non-local transform. At the linearized level and in 
momentum-space these are given
simply by momentum dependent leg pole factors \cite{Natsuume:1994sp,
Polchinski:1994mb}. This non-locality implies a
fuzziness of our causal diagrams at the {\em string scale} 
(at least in the asymptotic region). This fuzziness
is present in any description of spacetime in string theory. This does not,
therefore, modify our conclusions about the casual structure which is
in any case a concept at distance scales larger than the string length.
\footnote{In the nonlinear theory the transform is both non-local and
  nonlinear \cite{Dhar:1995hm}. 
This nonlinearity is not relevant for the considerations
  of this paper.}

We complete our analysis of the spacetime background
by  computing the form of the tachyon condensate
in the bulk of the weakly coupled region.  To this end, 
we compute the macroscopic loop transform of our solutions,
effectively probing the string spacetime with the end
of the FZZT brane \cite{Fateev:2000ik,Teschner:1995yf}.  
We find that the asymptotic behavior
of the tachyon profile so determined matches our expectations.

Appearance of spacelike ${\cal I}^\pm$ is associated with the
existence of cosmological horizons, and 
is reminiscent of de Sitter spacetimes.  Perhaps an in-depth study of these
scenarios could shed some light on the quantum mechanics
of de Sitter.

This paper is organized as follows: in section \ref{sec:solns}
we describe our class of time dependent solutions to the
matrix model and obtain the spacetime causal structure for these solutions,
represented by the Penrose diagrams.  
In section \ref{sec:mapping}, we discuss the
non-locality in the matrix model-to-spacetime map
and some of its consequences, introducing both
the leg pole transform and the macroscopic loop transform.
In  \ref{sec:loop} we discuss the macroscopic 
loop transform in detail and motivate its use for our purpose.
Finally, in  \ref{sec:tachyon}, we apply the
loop transform to obtain information about the
tachyon condensate.

\section{Moving Fermi sea solutions}
\label{sec:solns}

The classical limit of the matrix model is described
by motion of an incompressible Fermi fluid in phase
space under an equation of motion imposed by a Hamiltonian
\be
H = \half (p^2 - x^2)~.
\ee
We will use $\alpha'=1$ when discussing bosonic string
theory, and $\alpha'=\half$ when discussing 0B theory,
which allows us to use the same matrix Hamiltonian
for both cases.

The static Fermi sea profile, given by $(x-p)(x+p) = 2\mu$, corresponds
to a flat linear dilaton background with a tachyon wall whose position
is specified by the value of string coupling at the wall, 
$g_s \sim \mu^{-1}$.  As is well known, the
effective field $\eta$ describing small fluctuations about the profile
is massless and one can define two coordinate patches in the two
regions $x^2 > 2\mu$, each with
coordinates $\sigma$ and $\tau$,
 by the relations $x = \pm 2\mu \cosh \sigma$ (the $\pm$
refer to the two sides of the potential)
and $\tau = t$, in which the quadratic action is simply
\be
\label{eqn:S}
S = {1\over 2}\int d\sigma d\tau ( (\partial_\tau \eta)^2
-(\partial_\sigma \eta)^2 )~.
\ee
Up to string scale non-localities to be described later, the string
theory spacetime is closely related to the spacetime defined by
$\sigma$ and $\tau$.  The metric inferred from (\ref{eqn:S}) is of the
form $\exp(\rho) \eta_{\mu\nu}$ in the $\sigma$, $\tau$ coordinates.
The conformal factor $\rho$ cannot be determined from this analysis,
but the conformal structure can: it is simply half of flat space.  At
$\sigma=0$, there is a reflecting boundary condition, since
$x(-\sigma) = x(\sigma)$.  This corresponds to the tachyon wall in
spacetime. Interactions of the $\eta$ fields are cubic at the
classical level when expressed in terms of $\eta$ and its canonical
conjugate $\Pi_\eta$ and the coupling constant is $g(\sigma) = 1/(2\mu
\sinh^2 \sigma)$, which shows that the theory is strongly nonlinear
near the mirror.

The static Fermi sea is, of course, not the only solution
of the equations of motion. A class of nontrivial 
solutions can be obtained
by acting with various $W_\infty$ transformations on this
static profile \cite{Das:2004hw}. 
In phase space, one class of transformations 
is given by
\ben
(x \pm p) \rightarrow (x \pm p) + \lambda_\pm~e^{\pm rt}~(x\mp p)^{r-1} ~,
\label{eq:one}
\een
where $r$ is a non-negative integer and 
$\lambda_\pm$ are finite parameters. This leads to a class of 
time dependent Fermi surfaces
\ben
x^2-p^2 + \lambda_-~e^{-rt}(x+p)^r + \lambda_+ ~e^{rt} (x-p)^r +
\lambda_+\lambda_- (x^2-p^2)^{r-1} = 2\mu~.
\label{eq:two}
\een
For $r=1,2$ the profile of the Fermi surface in phase space is quadratic 
(ie, it is intersected by a constant $x$ line at most twice) and 
therefore corresponds to a classical solution of the collective field theory 
\cite{Dhar:1992cs, Das:2004rx}.
Non-quadratic profiles generically signify large quantum fluctuations
\cite{Das:1995gd,Das:2004rx}. 

The case of $r=1$ was studied in some detail in 
\cite{Karczmarek:2004ph}-\cite{Karczmarek:2004yc}.  
Here, we will focus on $r=2$.

Special coordinates, defined up to a conformal transformation
as the coordinates for which the quadratic action for small fluctuations
is of the form given in equation (\ref{eqn:S}), can be found for non-static
Fermi sea profiles as well \cite{Alexandrov:2003uh,Ernebjerg:2004ut}.
We will refer to these special coordinates as Alexandrov coordinates,
and use them to define the causal structure of the theory.
The philosophy here is that the collective field of the matrix
model corresponds to an effective massless field in spacetime,
and can therefore be used as a spacetime probe.  However, since
in two spacetime dimensions the kinetic term for such a field does 
not depend on the conformal factor of the metric, this probe does
does not allow us to determine this factor.

\subsection{The closing hyperbola solution}

\begin{figure}[ht]
   \vspace{0.5cm}

\centerline{
   {\epsffile{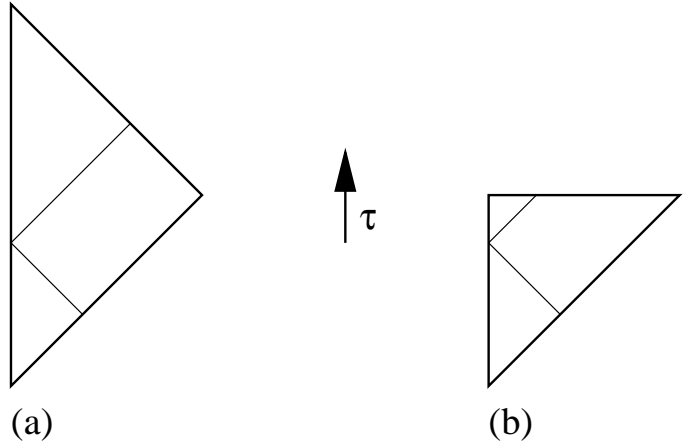}}
}
\caption{Causal structure of (a) Static flat space and (b) space
resulting from the solution in equation (\ref{eqn:closing}).
A sample null trajectory is shown ending on ${\cal I}^+$.}
\label{causal.intro}
\end{figure}

We begin with a special case of equation (\ref{eq:two}),
with $r=2$, $\lambda_- = 0$, and $\lambda_+ < 0$. By choosing the
origin of $t$, we may choose $\lambda_+ = -1$.
Hence, consider the following profile (in this
example, we will focus on just the right branch of the hyperbola):
\be
(x-p)(x+p + e^{2 t} (p-x)) = 2\mu~.
\label{eqn:closing}
\ee
This represents a hyperbola which starts out near the static configuration,
but eventually `closes' and escapes to infinity.  The Alexandrov coordinates
are given by the following coordinate transformation
\bear
x &=& \sqrt{2\mu} \frac {\cosh\sigma} {\sqrt{1-e^{2\tau}}}~, \\
t &=& \tau - \half \ln \left (1- e^{2\tau}  \right)~,
\label{coord:closing}
\eear
where the coordinate patch $\sigma>0$ and $\tau<0$ is enough to
cover the {\it entire} evolution of the Fermi surface. 
 Penrose diagram of this
spacetime, which exhibits its causal structure, is shown
in figure \ref{causal.intro}(b).   Though this example is very
simple, it is nontrivial, since ${\cal I}^+$ is
spacelike.

\begin{figure}[ht]
   \vspace{0.5cm}
\centerline{
   {
   \epsffile{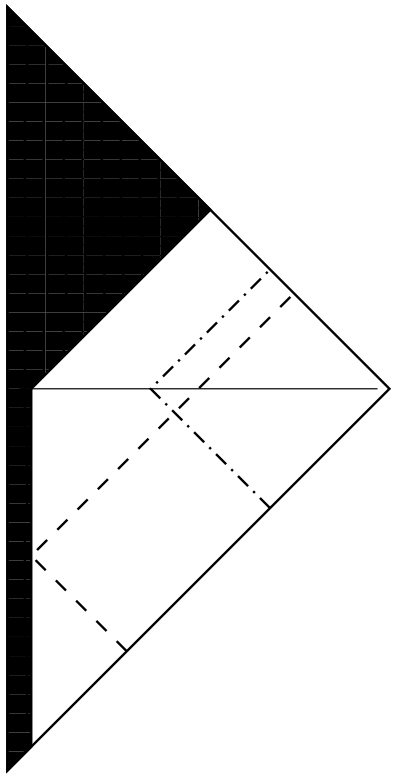}}
}
\caption{Trajectories of massless particles in fermion
coordinates $(t,q\equiv\log x)$ corresponding to spacetime
structure in figure \ref{causal.intro}(b).}
\label{fig3}
\end{figure}

This space-time is in fact geodesically incomplete and normally one would
of course extend this to full (half) Minkowski space. However, in terms of
the original matrix model time, this would mean that one has to extend
beyond $t = \infty$, which does not make sense. The underlying matrix model
therefore forces us to have this space-like boundary, perhaps suggesting
that the spacetime effective theory is strongly coupled there.

The causal structure should be compared to the spacetime
traced out by trajectories of points on the Fermi surface.  These are a one-parameter family of curves
\be
x_0(t) = {\sqrt{2\mu}}[ e^t \cosh \tau_0 + {1\over 2}e^{\tau_0 - t}]~,
\label{eqn:aone}
\ee
where $\tau_0$ is a real parameter. This is an {\em exact} solution of the
equations of motion and therefore gives the motion of the center of an
infinitesimal pulse is localized at $x_0$. 

Interestingly these curves are {\em null rays} in Alexandrov
coordinates.  The ray is given by $\tau + \sigma =\tau_0$ for an
incoming pulse and $\tau - \sigma = \tau_0$ for the reflected pulse,
as may be verified by substituting (\ref{eqn:aone}) in
(\ref{coord:closing}). Null rays are the trajectories of centers of
wave packets at the {\em linearized} level since the fluctuations are
massless scalars. It might appear strange this this continues to be
the case in the full nonlinear theory. However this simply follows
from the equation of motion. The equation ${\ddot x} = x$ written out in
Alexandrov coordinates (\ref{coord:closing}) becomes
\ben
\cosh \sigma[e^{4\tau} - 2 e^{2\tau}] 
+ {\dot{\sigma}}^2~\cosh \sigma [2e^{2\tau}-e^{4\tau}-1]
+ {\ddot{\sigma}}~\sinh\sigma[2e^{2\tau}-e^{4\tau}-1] = -\cosh
\sigma~,
\een
where dot denotes a derivative with respect to the Alexandrov time $\tau$.
It is clear from this that null rays ${\dot{\sigma}}=\pm 1$
automatically solve this equation.

In figure \ref{fig3} we show these trajectories in roughly 
the fermion coordinates, $t$ and $q \equiv \log x$.
Notice that this is {\it not} a Penrose diagram, as 45 degree
lines are not necessarily null.

\subsection{The opening hyperbola solution}

\begin{figure}[ht]
   \vspace{0.5cm}
\centerline{
   {\epsffile{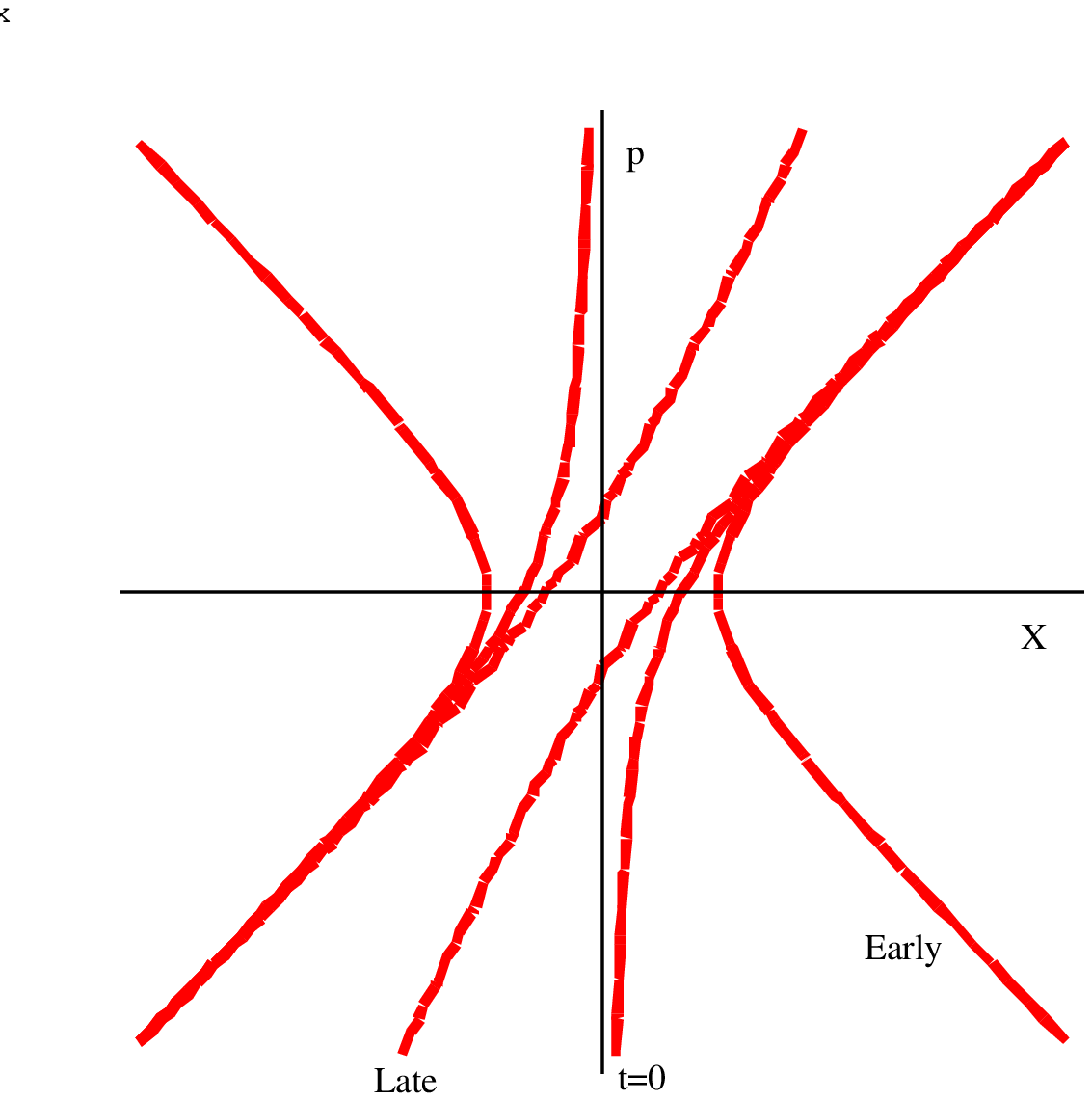}}
}
\caption{Time evolution of the Fermi sea given by equation
(\ref{eqn:opening}).}
\label{fermi}
\end{figure}

We will now focus on the following solution:
\be
(x-p)(x+p + e^{2 t} (x-p)) = 2\mu~.
\label{eqn:opening}
\ee
In terms of equation (\ref{eq:two}) this corresponds to $r=2$
with $\lambda_- =0$ and $\lambda_+ > 0$. We have further chosen the zero of time $t$ to
set $\lambda_+ = 1$.
The solution represents two branches of a hyperbola which approach the
static solution at $t \rightarrow -\infty$ and which then
`open up' and spill over the top of the potential at $t=0$,
see figure \ref{fermi}.
 
The configuration is qualitatively different for $t<0$ and $t>0$.
For $t<0$, each branch of the hyperbola is on its own side of the
potential, and each branch intersects a vertical line twice.  
The eigenvalue density $\varphi_0 (x,t)$ may be easily found to be
\be 
\varphi_0(x,t)  = 
{\sqrt{x^2 - 2\mu (1-e^{2t})}\over 1-e^{2t}}~.
\label{threetwo}
\ee
This clearly shows that for $t < 0$ the density of eigenvalues vanishes in the
region
\ben
|x| < {\sqrt{2\mu(1-e^{2t})}}~,
\een
whereas for $t > 0$ we will see there is no such cut.

For $t < 0$, the  techniques for defining Alexandrov coordinates
used in \cite{Alexandrov:2003uh} can be applied.  
The result is ($\pm$ correspond to the
right and left branch)
\bear
x &=& \pm \sqrt{2\mu} \frac {\cosh\sigma} {\sqrt{1+e^{2\tau}}}~, \nn \\
t &=& \tau - \half \ln \left ( 1+e^{2\tau} \right)~.
\label{eq:three}
\eear
These Alexandrov coordinates consist of two patches (corresponding to
the two branches of hyperbola), each covering the region $\sigma>0$ and 
$-\infty<\tau<\infty$.  The cut in eigenvalue distribution is at (in
fermion coordinates $x$) 
\be
x_{end}^{\pm} = \pm \sqrt {2\mu(1+e^{2\tau})^{-1}}~.
\ee
As $t \rightarrow 0$, $\tau \rightarrow \infty$, 
$x_{end}^{\pm}\rightarrow 0$, and the left branch and right
branch cuts meet.  This corresponds to the
hyperbola becoming `vertical' (see figure \ref{fermi}).

At $t=0$, the mirror disappears, and the eigenvalue 
density as a function of $x$ becomes infinite.  
For $t>0$, then, the effective field must be defined a little
differently. The two branches of the solution are now
`horizontal', that is, $p(x)$ is a single-valued function of $x$
on each branch.  We parametrize the two branches via
\bear
x_{\pm}(\omega,t) &=& 
\pm\sqrt{2\mu} \left (\cosh \omega - \half e^{2t - \omega} \right
)~,\nn \\
p_\pm(\omega,t) &=& 
\pm \sqrt{2\mu} \left (\sinh \omega - \half e^{2t - \omega} \right )~.
\label{eq:four}
\eear
and define the collective field as $\varphi_0 = p_-(x) - p_+(x)$.  
This is
\be
\varphi_0(x,t)  = -
{\sqrt{x^2 + 2\mu (e^{2t}-1)}\over e^{2t}-1}~.
\label{three2}
\ee
Notice that this is the same expression as for $t<0$, though
the interpretation is different.  (\ref{threetwo}) is the actual
eigenvalue density, while (\ref{three2}) represent the `eigenvalue
density minus infinity', or the negative of the density
of the complement of the eigenvalue distribution.

With these definitions,
an approach similar to that of Alexandrov can now be followed, 
defining $\tilde \omega(\omega,t)$ 
such that $x_-(\tilde \omega,t) = x_+(\omega,t)$, and then
the Alexandrov coordinates $\tau = -t+(\omega+\tilde\omega)/2$ and
$\sigma = (\omega - \tilde\omega)/2$. 
\bear
x &=& \sqrt{2\mu} \frac {\sinh\sigma} {\sqrt{e^{-2\tau}-1}}~, \nn \\
t &=& -\tau - \half \ln \left ( e^{-2\tau} -1 \right)~,
\label{eq:five}
\eear
is then the desired coordinate change.
This patch of Alexandrov coordinates has $\tau<0$ and 
$-\infty<\sigma<\infty$.

The crucial fact about this coordinate transformation is that the
endpoint of the fermion time evolution, $t = \infty$, corresponds to
$\tau = 0$. The underlying fermion dynamics instructs us to truncate
the spacetime at this spacelike surface.

\begin{figure}[ht]
   \vspace{0.5cm}
\centerline{
   {\epsffile{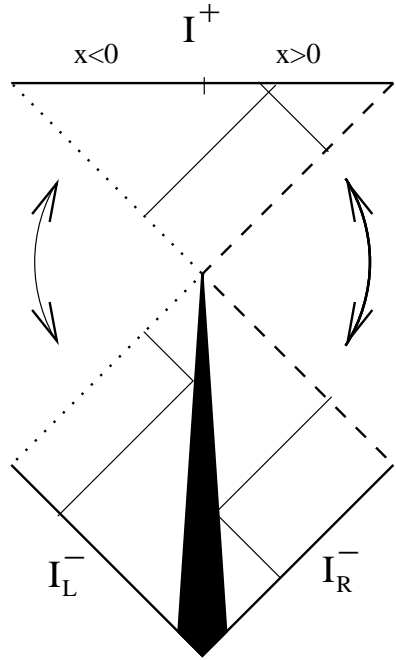}}
}
\caption{Causal structure obtained for the solution (\ref{eqn:opening}).}
\label{causal}
\end{figure}

We need to make sure that these are the correct coordinate 
transformations.
From the general formulae in \cite{Das:2004hw} (equation (36)) it
follows that the fluctuations are massless particles in a metric which
is conformally equivalent to
\ben
ds^2 = {1\over (1-e^{2t})^2} \left (
-dt^2+ {\left[(1-e^{2t})dx+e^{2t}x~dt\right]^2 \over
    x^2 - 2\mu(1-e^{2t})} \right  )~.
\label{threethree}
\een
In this metric, $t=0$ is a coordinate singularity. The
transformations (\ref{eq:three}) and (\ref{eq:five}) render the
metric Minkowskian 
\ben
ds^2 = -d\tau^2 + d\sigma^2
\een
in each patch.

The trajectory of a point on the Fermi surface $x(t)$ may be easily
determined and turns out to be
\ben
x(t) = \pm {\sqrt{2\mu}}
\left ( - e^t \sinh \tau_0 + {1\over 2}e^{-t} e^{\tau_0} \right )~.
\label{threeseven}
\een
Once again this is the {\em exact} trajectory. It is a null ray, as
may be seen by expressing this in Alexandrov coordinates. For $t < 0$
the incoming null rays are given by $\tau + \sigma = \tau_0$ while the
reflected ray is
given by $\tau + \sigma = \tau_0$. In the $t > 0$ region the
trajectory (\ref{threeseven}) corresponds to a null ray $\tau + \sigma
= -\tau_0$.

A particle which starts early on ${\cal I}^-_{R/L}$ (in our
conventions this means $\tau_0 < 0$) will end up on the
same side of the potential as the one on which it started.  
A particle which starts out later will  cross the $x=0$ line 
and end up on the other side of the potential. 
Every particle is reflected from the ``mirror'' at $\sigma = 0$
(independent of the value of $\tau_0$), but
this does not imply a reflection in $x$ space. 
In fact for trajectories with $\tau_0 > 0$, $x(t)$ is monotonic.

By following particle trajectories, we can glue the three coordinate
patches together to describe the causal structure of the entire
spacetime, as is shown in figure \ref{causal}.  The lower,
diamond-like patch corresponds to $t < 0$ while the upper triangle is
the region $t > 0$. The dashed and the dotted null lines are
identified as shown. The trajectory on the right side of the diagram
corresponds to an incoming pulse with $\tau_0 < 0$ while the
trajectory which starts on the left side is an incoming pulse on the
other side with $\tau_0 > 0$.  The causal structure is quite
intricate, and spacetime ends with a spacelike boundary ${\cal I}^+$
which is $\tau =0$ of the $t > 0$ patch.

Once again one would normally extend the space-time beyond this
space-like boundary, but the fundmental description in terms of the
matrix model makes this extension meaningless.

\begin{figure}[ht]
   \vspace{0.5cm}
\centerline{
   {\epsfxsize=8.5cm
   \epsfysize=8cm
   \epsffile{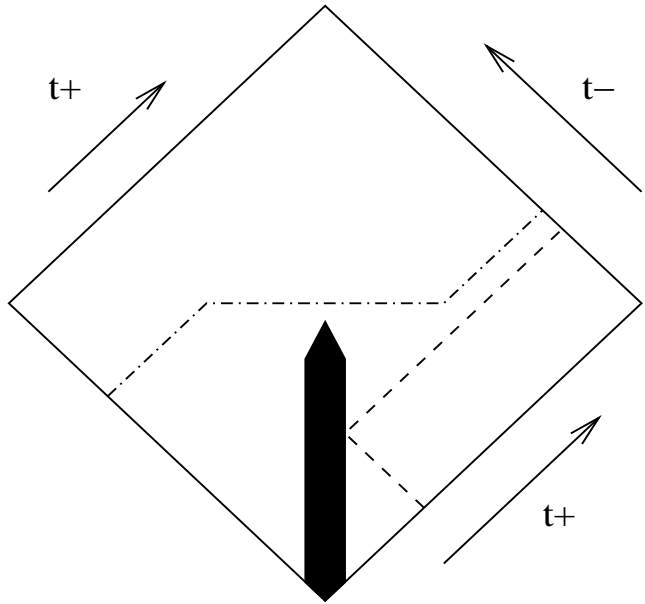}}
}
\caption{Diagram in fermion $t^\pm$ coordinates, for solution
(\ref{eqn:opening}), with trajectories of massless particles
shown.}
\label{fig1}
\end{figure}

These null trajectories are also shown in figure \ref{fig1}, in fermion
coordinates. Once again, this is not a Penrose diagram.

The diagrams in both figure \ref{causal} and in
figure \ref{fig1} have to be folded along the central
vertical line since in the type 0B interpretation 
\cite{Douglas:2003up,Takayanagi:2003sm}
both sides of the potential correspond to the same spacetime region . 
It is therefore probably best to interpret figures
\ref{causal} and \ref{fig1} as representing a double cover
of spacetime, with symmetric and antisymmetric fluctuations
corresponding to the two spacetime fields of 0B theory.

For simplicity, we have so far restricted our 
attention to solutions which asymptotically approach the
ground state in the past ($\lambda_- = 0$).
This lead us to a spacelike ${\cal I}^+$ and a null ${\cal I}^-$. 
It is evident that a parallel  discussion 
for the time reversed version would yield a
spacelike ${\cal I}^-$ and null ${\cal I}^+$. 
It is also possible to take both $\lambda_\pm \neq 0$ 
in equation (\ref{eq:two}).
We will not work out the details of the spacetime structure, but
general features are obvious: in this
case both ${\cal I}^+$ and ${\cal I}^-$ are spacelike.

\section{Mapping to string theory spacetime}
\label{sec:mapping}

It is well known that the complete dictionary between the matrix model
and string theory in spacetime involves a non-local transformation.  In
momentum space these are the "leg-pole factors".
The S-matrix obtained from the matrix model has to be
multiplied by these additional momentum dependent phases to match
results from the string theory world-sheet.  In space, these momentum
dependent factors result in a non-local transform, which relates the
matrix model effective field $\eta$ to the spacetime string field
$S(\phi, t) = e^{-\Phi}T(\phi, t)$, where $T$ is the tachyon.  Since
this transform is a statement about the S-matrix, it has meaning only
on ${\cal I}^\pm$.  On ${\cal I}^+$ it is given by
\be
S(t-\phi) = \int dv K(v) \eta(t-\phi-v)~.
\ee
In this section we wish to investigate whether this non-locality modifies
our causal structure.

\subsection{Leg Pole Kernels} 
For fluctuations around the usual ground state the leg-pole kernel is
known both for the bosonic and the 0B theory.  
In the bosonic string theory the kernel $K$ is given by 
\cite{Polchinski:1994mb}
\bear
K_{bos}(v) &=& \int_{-\infty}^{\infty} {d\omega \over 2\pi} 
e^{i\omega v}
\left
(\pi / 2 \right )^{-i\omega/4} {\Gamma(-i\omega) \over
\Gamma(i\omega)} = - {z \over 2} J_1 (z) = {d \over dv}
J_0(z)~, \\ \nn
z(v)&\equiv& 2 (2/\pi)^{1/8} e^{v/2} ~.
\label{legpoleK}
\eear
We will denote the Fourier transform of the kernel by
$L_B(\omega)$, 
\be
L_{B}(\omega) =\left
(\pi / 2 \right )^{-i\omega/4}
 \Gamma(-i\omega)/\Gamma(i\omega)~.
\label{legpoleboson}
\ee
 The asymptotic behavior  of the kernel is 
\be
K_{bos}(v) \sim e^v~,~~~~~ v \rightarrow-\infty 
\ee 
and 
\be
K_{bos}(v) \sim e^{v/4} \cos(z + \pi/4)
~,~~~~~ v \rightarrow +\infty ~.
\ee
$K_{bos}$ decays exponentially for $v \rightarrow -\infty$ and grows 
while oscillating wildly for $v \rightarrow +\infty$.

The precise form for this leg pole transform is different in the 0B theory.
Here we have two massless scalar fields in spacetime : the tachyon
$T_{NSNS}$ and the axion $C_{RR}$ living in the {\em same}
spacetime. There are two fields in the matrix model as well --- these
are the fluctuations of the Fermi sea on the two sides of the potential.
In terms of the collective field fluctuation $\eta (x,t)$, written as a
function of the ``fermion'' coordinate $x$, one can define symmetric
and antisymmetric combinations $\eta_{S,A}(x,t)$ which may be thought
to live on half of $x$ space, which we will take to be $x > 0$
\ben
\eta_{S,A}(x,t) = {1\over 2}[\eta(x,t) \pm \eta(-x,t)]~.
\een
For on shell fields the Fourier transforms of $\eta_{S,A}$ are
related to the spacetime fields by 
\cite{Douglas:2003up,Takayanagi:2003sm}
\bea
T_{NSNS}(\omega)  =  L_{NSNS}(\omega) ~\eta_S(\omega) &,& ~ 
L_{NSNS}(\omega) = \left
(\pi / 2 \right )^{-i\omega/8}
{\Gamma(i\omega / 2 )\over
  \Gamma(-i\omega /2)}~,
\label{eqn:legpoleNSNS}
\\
C_{RR}(\omega)  =  
L_{RR}(\omega)~\eta_A(\omega) &,& ~ 
 L_{RR}(\omega) = \left
(\pi / 2 \right )^{-i\omega/8}
{\Gamma\left({1\over 2}+{i\omega\over 2}\right)\over
  \Gamma\left({1\over 2} -{i\omega\over 2}\right)}~.
\label{eqn:legpoleRR}
\eea
This implies that the corresponding kernels are given by
\be
K_{NSNS}(v) = \int_{-\infty}^{\infty} {d\omega \over 2\pi} 
e^{i\omega v}
\left
(\pi / 2 \right )^{-i\omega/8} 
{\Gamma\left({i\omega\over 2}\right)\over
  \Gamma\left(-{i\omega\over 2}\right)}
= - z J_1 (z)
\ee
and
\be
K_{RR}(v) = \int_{-\infty}^{\infty} {d\omega \over 2\pi} 
e^{i\omega v}
\left
(\pi / 2 \right )^{-i\omega/8} 
{\Gamma\left({1\over 2}+{i\omega\over 2}\right)\over
  \Gamma\left({1\over 2} -{i\omega\over 2}\right)}
=  z J_0 (z)~,
\ee
where
\be
z(v) \equiv 2 (2/\pi)^{1/8} e^{v}~.
\ee
The asymptotic behavior for $v \rightarrow -\infty$ is in this case
\bear
\label{eqn:KRR}
K_{NSNS}(v) &\sim& e^{2v}~,
\nn\\
K_{RR}(v) &\sim& e^{v}~.
\eear
Recall that our conventions are $\alpha'=1$ in bosonic theory, and
$\alpha' = \half$ in 0B theory.  We are therefore using string units;
the formulae above show that the scale of the relative non-locality
between matrix model quantities and spacetime string fields is of the
order of the string length.

\subsection{Macroscopic loops}

Unfortunately, the leg pole transform gives no insight into what is happening
in the bulk of space, as it describes only the null infinities.  In
order to achieve some amount of insight into the bulk, we will resort
to the macroscopic loop transform. As discussed in detail in the next 
section, the macroscopic loop will define the notion of spacetime 
as perceived by the FZZT branes.

In the bosonic theory the macroscopic loop is defined by
\be
W(l,t) \equiv tr\Big(e^{-lM(t)}\Big) = \int_0^\infty
dx \varphi(x,t) e^{-lx}~,
\label{eqn:bosonic.loopW}
\ee
where $\varphi$ denotes the eigenvalue density and $l=e^{-\phi}/\sqrt{2}$.
This is again a non-local transformation on the collective field which is of course different from the leg pole transform. 
However, the two are approximately the
same in the asymptotic regimes ${\cal I}^\pm$.  Rewriting the
macroscopic loop transform for a small oscillation $\eta$ on top of
whatever background is under consideration, we get
\be
w(t,\phi) = \int dx \exp(- {1\over {\sqrt{2}}}e^{-\phi} x) 
\partial_x \eta=
\int dq \exp\big(-\sqrt{\mu} e^{-\phi} \cosh q \big)\partial_q \eta~.
\ee
Consider the outgoing modes, with
$\phi \rightarrow +\infty$ and a finite support for the wave-packet
$w(t-\phi)$.  Then 
\be
\label{loopout}
w(t-\phi) = 
\int dy \left[e^y \exp\big(-e^y \big) \right ]
\eta(t-\phi-y-\log({{\sqrt{\mu}}\over 2}))~.
\ee 
To compare with the leg pole transform we need to shift the origin of
time by an amount $\log({{\sqrt{\mu}}\over 2})$.  Comparing with the
kernel which appears in square brackets in (\ref{loopout}) with the
leg pole kernel $K$, we notice that the two kernels have the same
exponential tail for large negative argument.  The macroscopic loop
transform approximates the exact leg pole expressions here. The two
transforms start to differ when the argument is close to zero: for
positive values of the argument $K$ has wild oscillations while kernel
in (\ref{loopout}) decays very rapidly.  The Fourier transform of the
macroscopic loop kernel is
\be
\int dy e^{-i\omega y} \sqrt\mu e^y \exp\big(-\half \sqrt\mu e^y \big)
\sim
\Gamma(i\omega) ~.
\ee
Comparing with (\ref{legpoleK})
we see that the two kernels have precisely the same
poles, but the residue is the same only for the first pole, which
determines long range behavior.

In the 0B theory, the macroscopic loop transform needs to
be modified appropriately \cite{Douglas:2003up, Takayanagi:2003sm}.  
The NSNS loop is defined by
\be
W_{NSNS}(l,t) = Tr e^{-l(M(t)^2-\mu)} =  \int_{-\infty}^{\infty}
dx ~e^{-l(x^2 - \mu)} \varphi (x,t)~,
\label{loop:NSNS}
\ee
where now $l=e^{-2\phi}$.
The fluctuation of this quantity is given in terms of the
symmetric field $\eta_S$ introduced above 
\be
w_{NSNS}(\phi,t) = 2 \int_0^\infty d\sigma e^{-e^{-2\phi}\mu \cosh(2\sigma)}
\partial_\sigma \eta_S~.
\ee
The kernel of this transformation is consistent with the $e^{2v}$ fall-off
of the NSNS leg-pole transform.

In the RR sector there are two macroscopic loops (we are
ignoring here the GSO projection)
\be
W^\pm_{RR}(l,t) = \sqrt{2l}~
Tr \left ( M(t)  e^{-l(M(t)^2 \mp\mu)}  \right ) = \int_{-\infty}^\infty  
dx ~\sqrt{2l}~x~e^{-l(x^2 \mp \mu)} \varphi (x,t)~.
\label{loop:RR}
\ee
The fluctuation of $W^+_{RR}$ is given by
\be
w^+_{RR}(\phi,t) = 2 \int_0^\infty d\sigma \sqrt{2\mu} e^{-\phi}
\cosh{\sigma} e^{-e^{-2\phi}\mu \cosh(2\sigma)}
\partial_\sigma \eta_A~.
\ee
The kernel falls of like $e^\phi$, again in
agreement with (\ref{eqn:KRR}).  Similar statements
can be made about $w^-_{RR}$.

\subsection{Implications for time-dependent solutions}

The leg pole factors were derived by comparing the S-matrix of
perturbative fluctuations around the ground state
obtained from matrix model with the S-matrix obtained from world-sheet
string theory. In general time dependent backgrounds the world-sheet
formulation is not known at the moment and a general principle which
determines these transforms is lacking. In the absence of such a
principle we assume that the generic features of the correct
transform would be the same as around the ground state.

In the previous sections we obtained the causal structure of the
spacetime generated by our time dependent solution in Alexandrov
coordinates, which are locally related to the matrix model coordinates
$(x,t)$. Since the spacetime perceived by perturbative strings is
different from this spacetime and related to it by a non-local
transform, sharp null rays in the Penrose diagrams would correspond
spread out pulses in the perturbative string spacetime. Assuming
that the scale of non-locality introduced by a leg pole transform
is still the string scale, one has to conclude that the diagrams are
smeared out, but only by an amount which is of the order of the string
length. Probes which have low energies will continue to perceive the
spacetime exactly as depicted.

This may be checked by an explicit calculation of the macroscopic
loop, whose arguments provide spacetime as perceived by FZZT branes.
Consider for
example a fluctuation which has the form
\ben
\eta (x,t) = {1\over \sigma {\sqrt{\pi}}}~{\rm exp}~[-{(x-x_0(t))^2
    \over \sigma^2}]
\een
for some trajectory $x_0(t)$ of its center.
Then the corresponding NSNS macroscopic loop is
\ben
W(l) = {2l x_0 (t) \over (l\sigma^2 +1)^{3/2}}~{\rm exp}~
[-{l x_0^2(t) \over l\sigma^2 +1}]
\een
In the asymptotic region $l \rightarrow 0$ it is clear that $W(l)$
peaks at $l = {1\over x_0^2(t)}$ with a width in $l$ space of the same
order. This means that in the $\phi$ space the width is of order
unity. Recalling that in our discussion all lengths are measured in
units of the string scale we therefore see that the width in physical
$\phi$ space (which agrees with the space of string theory in the
asymptotic region) 
is of the order of
the string length\footnote{The same result holds for the bosonic
theory as well.}. In the 0B interpretation one has to fold over the
Penrose diagram in (\ref{causal}) along $x=0$ with a proper
identification of the null rays as appropriate mixtures of the tachyon
and the axion field. Our discussion shows that the causal structure
Penrose diagrams in figures \ref{causal.intro} and \ref{causal}
reflect quite accurately the
string spacetime $\phi-\tau$ in the region of large $\phi$.

\section{Tachyon condensates}

Consider perturbing the Liouville world-sheet action by adding a tachyon
condensate:
\be
S = S_{Lioville} + {1 \over 4\pi} \int dz^2 T(\phi,t)~.
\ee
We are interested in the relationship between the 
tachyon condensate and the perturbed solutions of the matrix
model from section \ref{sec:solns}.  One way to probe
such a perturbation of the world-sheet action is via a 
one point function: when the one point function is computed
in perturbation theory for a small perturbation of the
world-sheet action, the first nontrivial term is proportional
to the tachyon condensate.
At the same time, the one point function on the 
string world-sheet corresponds to a wave functional, and so
it should satisfy a minisuperspace wave equation.
It was shown in \cite{Fateev:2000ik} that the Laplace
transform of the FZZT disc one point function satisfies the
minisuperspace wave equation exactly, and not just in the classical
(minisuperspace) limit.
In this section, we will propose to use the Laplace transform
of the one point function, which is just the macroscopic loop
introduced above, to obtain a candidate for a world-sheet
perturbation $T(\phi,t)$.  

\subsection{Macroscopic Loops and FZZT branes}
\label{sec:loop}

We will review here the salient facts from 
\cite{Douglas:2003up, Takayanagi:2003sm,Takayanagi:2004ge}.

The macroscopic loop is related via a Laplace transform to a 
disc one-point function with an FZZT brane boundary.  The FZZT branes
\cite{Fateev:2000ik,Teschner:1995yf} 
are labeled by a boundary cosmological constant $\mu_B$.
They should be thought of as D1-branes extending from the weak coupling
region toward the strong coupling region, and dissolving at
a point determined by $\mu_B$.  These branes provide
quasi-local probes of the geometry of spacetime generated by the 
matrix model \cite{Martinec:2003ka,Seiberg:2003nm}.  
The non-locality can be removed by taking a Laplace transform, 
at which point we arrive at the macroscopic loop, which is defined
by cutting a fixed-length hole in the world-sheet, and thus
holding the boundary at a constant dilaton value.

Consider therefore a FZZT brane with a boundary
cosmological constant $\mu_B$ in the bulk background parametrized
by the bulk cosmological constant $\mu$.  
In bosonic theory, up to a normalization constant, 
the one point function is \cite{Fateev:2000ik,Teschner:1995yf} 
\footnote{We are quoting
these formulae for a general quantum improvement term $Q = b+1/b$,
in order to regularize the expressions.}
\be
\label{eqn:onepoint}
\langle V_{ik} \rangle_{FZZT} \sim
{\cos(\pi s k) \over i k} \Gamma(1+ibk)\Gamma(1+ib^{-1}k)~,
\ee
where the parameter $s$ is related to $\mu_B$ via
\be
\cosh^2 (\pi b s) = {\mu_B^2 \over \mu} \sin(\pi b^2)~. 
\ee
For $b=1$, we can rewrite (\ref{eqn:onepoint}) as
\be
\langle V_{ik} \rangle_{FZZT} \sim
\cos(\pi s k) \Gamma(1+ik)\Gamma(ik)~.
\label{eqn:onepoint.b=1}
\ee
This quantity is related via a Laplace transform to the
macroscopic loop defined in (\ref{eqn:bosonic.loopW}).
The relationship is as follows: consider
a perturbation of the eigenvalue density $\varphi(x,t)
= \sqrt{x^2- 2\mu} +  \partial_x \eta$.  The macroscopic loop
of the perturbation alone is then $(x=\sqrt{2\mu} \cosh\sigma)$
\be
w(l,t) = \int_0^\infty d\sigma \partial_\sigma \eta(\sigma,t)
e^{-l\sqrt{2\mu}\cosh\sigma}~.
\label{eqn:bosonic.loopw}
\ee
From the discussion of Alexandrov coordinates in the
previous section we know that $\eta(\sigma, t=\tau)$ satisfies the standard massless Klein-Gordon equation $[\partial_\tau^2 - \partial_\sigma^2]\eta = 0$ at the linearized level.  Consider therefore a mode of this field
given by $\eta = e^{\pm ikt} \sin(k \sigma)$.
Substituting this into (\ref{eqn:bosonic.loopw}), we obtain
\be
 w(l,k) = \int_0^\infty d\sigma 
e^{-l\sqrt{2\mu}\cosh\sigma} k \cos(k\sigma) = 
k K_{ik}(\sqrt{2 \mu} l)~.
\ee
Taking a Laplace transform of this result, we obtain
\be
\int_0^\infty {dl \over l} e^{-\sqrt{2\mu}l\cosh(\pi s)} w(l,k) = 
{\pi \cos(\pi k s)\over \sinh(\pi k)}~.
\ee
To relate this to the worldsheet theory, we need to multiply by
the leg-pole transform in equation (\ref{legpoleboson}),
obtaining
\be
-i \cos(\pi k s) \Gamma(1+ik)\Gamma(ik)~,
\ee
which agrees with (\ref{eqn:onepoint.b=1}).
Furthermore, the fluctuation of a loop transform (\ref{eqn:bosonic.loopw})
satisfies the minisuperspace equation for a tachyon field, given by
\ben
[\partial_t^2 - (l\partial_l)^2 + 2\mu l^2] w(l,k) = 0~.
\label{mfoura}
\een

In the 0B theory,
the FZZT-brane one point functions have been computed in
\cite{Fukuda:2002bv}, and are given by 
\bear
\label{eqn:1}
\langle V^{NSNS}_{ik} \rangle_{FZZT,\pm} &\sim&
{\cos(\pi k s) \over ik} \Gamma(1+ikb/2)\Gamma(1+ib^{-1}k/2)~, \\
\label{eqn:2}
\langle V^{RR}_{ik} \rangle_{FZZT,+} &\sim&
{\cos(\pi k s) \over ik} \Gamma(1/2+ikb/2)\Gamma(1/2+ib^{-1}k/2)~, \\
\label{eqn:3}
\langle V^{RR}_{ik} \rangle_{FZZT,-} &\sim&
{\sin(\pi k s) \over ik} \Gamma(1/2+ikb/2)\Gamma(1/2+ib^{-1}k/2)~, 
\eear
where the $\pm$ refer to the two different FZZT boundary
conditions possible in 0B theory.  

The macroscopic loops in the 0B theory have been defined in 
(\ref{loop:NSNS}) and (\ref{loop:RR}).
In both cases, the loop parameter $l$ is now related to the 
spacetime coordinate $\phi$ by $ l = e^{- 2 \phi}$.

Repeating the steps for bosonic string, we compute the
fluctuation $w_{NSNS}$ of the macroscopic loop $W_{NSNS}$ due to a mode $\eta=\sin(k\sigma)$
with $x=\sqrt{2\mu} \cosh \sigma$, to obtain
\be
w_{NSNS}(l,k) = k K_{ik/2}(\mu l)~.
\ee
This quantity has to be multiplied by the leg pole factor for 
NSNS fields given in (\ref{eqn:legpoleNSNS}).
Computing the Laplace transform we get
\be
{\Gamma\left({ik\over 2}\right) \over \Gamma\left({-ik\over 2}\right)}
\int {dl \over l} e^{-\mu l \cosh(2\pi s)} w_{NSNS} = 
{-i} \cos(\pi s k) \Gamma(1+ik/2) \Gamma(ik/2) ~,
\ee
which is seen to agree with (\ref{eqn:1}).
It is easy to check that $w_{NSNS}$ satisfies the minisuperspace
equation of motion \cite{Douglas:2003up}
\be
[\partial_t^2 -4(l\partial_l)^2 + 4\mu^2 l^2]w_{NSNS} =  0~.
\ee
For the RR field, there are two transforms defined in (\ref{loop:RR}).
The fluctuation of these quantities are
\ben
w^\pm_{RR} = 
\int_{-\infty}^\infty  
dx ~\sqrt{2l}~x~e^{-l(x^2 \mp \mu)} \partial_x \eta (x,t)~. 
\een
These two functions satisfy the equations of motion \cite{Douglas:2003up}
\bear
[\partial_t^2 -4(l\partial_l)^2 \pm 4\mu l +  4\mu^2 l^2]w^{\pm}_{RR} &=&  
0~,\\ \nn
[\pm 2(l\partial_l) +2\mu l ]w^\pm_{RR} 
&=& i\partial_t W^{\mp}_{RR} ~.
\eear
Again, we can recover the appropriate one-point functions
(\ref{eqn:2}) and (\ref{eqn:3}) by computing a Laplace transform
of the fluctuation $w$.  We must use $x=\sqrt{2\mu} \cosh\sigma$
for the ``$+$'' case and  $x=\sqrt{2\mu} \sinh\sigma$
for the ``$-$'' case, and the fluctuation must be given
by $\eta = \sin{k\sigma}$ and $\cos{k\sigma}$ respectively.
We then obtain
\bear
w_{RR}^+ &=&  {k \over 2} \sqrt{\mu l} \left(  
K_{1/2+ik/2} + K_{1/2 - ik/2}
\right)(\mu l)
\\
w_{RR}^- &=&  {-ik \over 2} \sqrt{\mu l} \left(  
K_{1/2+ik/2} - K_{1/2 - ik/2}
\right)(\mu l)
~.
\eear
Multiplying by the leg pole factor $L_{RR}$ (\ref{eqn:legpoleRR})
and computing the Laplace transforms (there is a factor of $\sqrt{l\mu}$
coming from fermionic modes), we obtain
\be
{\Gamma\left(\half + {ik\over 2}\right) \over 
\Gamma\left(\half - {ik\over 2}\right)}
\int {dl \over l}  \sqrt{\mu l} \left [ {\cosh \atop \sinh} \right ]
(\pi s) e^{-\mu l \cosh(2\pi s)} w^{\pm}_{RR} = 
{-i} \left[{\cos\atop \sin}\right ](\pi s k) (\Gamma(1/2+ik/2))^2 
\ee
which is in agreement with (\ref{eqn:2}) and (\ref{eqn:3})
for $b=1$.

Notice that the difference between $W_{RR}^+$ and $W_{RR}^-$ is 
small in the asymptotic region, since for $\phi \rightarrow
\infty$, $l\rightarrow 0$, $e^{\pm \mu l} \rightarrow 1$.

\subsection{The tachyon condensate}
\label{sec:tachyon}

We will now use the above expressions for the macroscopic loop 
transform to compute perturbations of the Liouville action
corresponding to our time dependent solutions from
section \ref{sec:solns}.

We begin by analyzing, in bosonic string theory, the simpler solution
whose fermi sea is given by equation (\ref{eqn:closing}).
The eigenvalue density is given by
\be
\varphi = \frac{\sqrt{x^2-2\mu(1+e^{2t}) }} {1 + e^{2t}}~.
\ee
Applying the bosonic loop transform (\ref{eqn:bosonic.loopW}),
we obtain
\bear
W(\phi, t) &=& \int_{{\sqrt{2\mu(1+e^{2t})}}}^\infty dx ~{\rm exp}[{-e^{-\phi} x}]~ 
{\sqrt{x^2 - 2\mu\left(1+e^{2t}\right)} 
\over {1+e^{2t}}} \nn \\ &=&
\sqrt{2\mu \left ( 1+e^{2t} \right )^{-1}} e^\phi
K_1 \left(  \sqrt{2\mu \left ( 1+e^{2t} \right )} e^{-\phi} \right )~.
\eear
In order to obtain the spacetime tachyon field from this expression,
it is necessary to subtract the background $\mu=0$ static solution given
by $\varphi = |x|$, with transform
\be
\int_0^\infty dx x  e^{-e^{-\phi} x} = e^{2\phi}~.
\ee
The tachyon field, after being dressed with the dilaton is
\be
T(\phi,t) = 1 - \sqrt{2\mu \left ( 1+e^{2t} \right )^{-1}} e^{-\phi}
K_1 \left(  \sqrt{2\mu \left ( 1+e^{2t} \right )} e^{-\phi} \right )~.
\ee
Let's look at some asymptotic behaviors of $T$.  First, early time
$t \rightarrow -\infty$.  This should correspond to the static
Liouville background.  We obtain 
\be
 T(\phi, t \rightarrow -\infty) =
1 - \sqrt{2\mu} e^{-\phi} K_1 (\sqrt{2\mu} e^{-\phi})~.
\ee
Using the fact that for small $x$, 
\be
K_1(x) \sim {1 \over x} + {x \over 2}
 \left ( \ln {x \over 2} - const \right )~,
\ee
we get that for large $\phi \gg 0$,
\be
 T(\phi \rightarrow +\infty, t \rightarrow -\infty) =
\mu e^{-2 \phi} \left ( \phi + const \right )~,
\ee
which is precisely the expected Liouville potential.

In the region where $1+e^{2t} \ll e^{2\phi}$, we get
\be
T(\phi>>0, t < \phi) = {e^{2t} \over 1 + e^{2t}}
+ \mu e^{-2 \phi} \left ( \phi + const
- \ln \sqrt{1+e^{2t}}\right)~.
\ee
Notice that the time dependent terms became important
around $t=0$, which is the point shown in figure
\ref{fig3} where the `null' trajectory suddenly turns around.

In order to get a feeling for how this looks in
the conformally flat coordinates, lets make the
association 
\bear
e^{2\phi} &\sim& {e^{2\sigma} \over 1-e^{2\tau}}\\
{e^{2t} \over {1+e^{2t}}} &=& e^{2\tau}
\eear
which follows from the approximate locality of the
loop transform and equation (\ref{coord:closing}).
We obtain the first few terms for the tachyon condensate,
in the regime where $\exp(\sigma) \gg 1$
\be
T(\sigma>>0, \tau < 0) = e^{2\tau} +
e^{-2\sigma}(1-e^{2\tau})(\sigma + const)~.
\ee 
This has a few interesting features.  First, we see that 
there is a {\it spacelike} condensation of the tachyon;
the first term, independent of $\sigma$ starts out zero
in the past and becomes important around $\tau=0$,
just before the world ends on the spacelike boundary.  
Second, we see that the Liouville potential is still there.
Third, we notice that the timelike and spacelike CFTs are coupled
through the last term.

The analysis of the solution given by (\ref{eqn:opening}) is more
complicated.  Since the fermi sea spills over the top of the
potential, it is necessary to analyze this in the 0B string theory,
using the more complicated loop transforms (\ref{loop:NSNS}) and
(\ref{loop:RR}).

The eigenvalue density for our solution is
\be
\varphi_0 = \frac{\sqrt{x^2+2\mu(e^{2t}-1) }} {1-e^{2t}}~.
\ee
Since the solution is symmetric under $x \rightarrow -x$,
the RR loop transform $W_{RR}$ vanishes.  For the NSNS
field, the integrals can be computed using the following formulae
\bear
\label{eqn:int}
2 \int_1^\infty dz e^{-\alpha x^2} \sqrt{z^2-1} &=& 
{e^{-\alpha/2} \over 2} \left (K_1(\alpha/2) - K_0(\alpha/2) \right )
\\
2 \int_0^\infty dz e^{-\alpha x^2} \sqrt{z^2+1} &=& 
{e^{\alpha/2} \over 2} \left (K_1(\alpha/2) + K_0(\alpha/2) \right )
\eear
and the answer is,
\be
W_{NSNS}= \mu e^{\mu l e^{2t}}  \Re \left (
K_1(\mu l (1-e^{2t})) - K_0(\mu l (1-e^{2t}))
 \right) ~,
\label{eqn:Wnsns}
\ee
where we have used $\Re(K_1(-|x|))=-K_1(|x|)$ and 
$\Re(K_0(-|x|))=K_0(|x|)$
to write a single formula encompassing $t<0$ and $t>0$.
The asymptotic behavior for small $l$ is
\be
\Re \left (K_1(x) - K_0(x) \right) 
\sim 
{1\over x} + \ln{|x|\over 2} - const~.
\ee
Since the loop transform equations for 0B depend on $\mu$,
we cannot subtract from (\ref{eqn:Wnsns}) the vacuum $\mu=0$
expression as we did before.  We must instead subtract
the loop transform of the background with a cosmological
constant,
\ben
\int_{-\infty}^\infty dx~\sqrt{x^2 - 2\mu}~e^{-lx^2} = 
\mu  \left (K_1(\mu l) - K_0(\mu l)
 \right) ~.
\label{eqn:Wnsns0}
\een 
Subtracting, then, (\ref{eqn:Wnsns0}) from (\ref{eqn:Wnsns}), 
multiplying by $e^\Phi$,
and retaining only the leading non-vanishing terms as $\phi \rightarrow \infty$,
we obtain that the tachyon condensate is
\bear
\delta T(\phi >> 0, t<0)& \sim &{e^{2t} \over 1- e^{2t}} = e^{2\tau}~,
\\ 
\delta T(\phi >> 0, t>0) &\sim &{-e^{2t} \over e^{2t}-1} = -e^{-2\tau}~.
\eear
Notice that the static Liouville term is not present; it has
been subtracted off in (\ref{eqn:Wnsns0}).


The properties of this condensate are somewhat similar to the ``closing
hyperbola'' solution. In particular there is a spacelike condensation
of tachyons.  However, the condensate is infinite at $t=0$, corresponding to the
junction of the coordinate patches shown in figure \ref{causal}.  
This is a sign that something quite singular and strongly coupled
must be happening that this point.

\section{Discussion}

The most interesting feature of our solutions is that ${\cal I}^\pm$
are spacelike. Spacetimes with spacelike ${\cal I}^\pm$, e.g. de
Sitter spacetime, have particle
and event horizons perceived by timelike geodesics. For two
dimensional string theory perturbative fluctuations are
massless. However there are various kinds of D-branes which are
massive and it would be interesting to investigate whether such
D-brane probes perceive horizons. This could shed some light on
origins of thermality associated with cosmological horizons.

It must be emphasized that the conformal structures are given in 
figures \ref{causal.intro} and \ref{causal}. 
The metric there is conformally flat.
The non-local transform to macroscopic loops lead to a
string length fuzziness in this diagram, but that is a feature of any
such diagram in string theory. On the other hand, 
the tachyon condensate is naturally given in fermion coordinates, such
as were used in figures \ref{fig3} and \ref{fig1},
where the metric is very nontrivial (these are nonconformal deformations
of the Penrose diagrams).  The tachyon condensate computed in
section \ref{sec:tachyon} seems in agreement with our expectations.

\section{Acknowledgements}

We would like to thank 
J.~ Davis, F.~ Larsen, G.~ Mandal, D.~ Shih, N.~Seiberg
and T.~Takayanagi for
discussions. The work of S.R.D. is partially supported by 
a National Science Foundation grant PHY-0244811 and a  DOE contract
DE-FG-01-00ER45832. The work of J.L.K. is partially supported by a
DOE contract DE-FG-02-91ER40654.

\section{Note Added} The results of this paper were briefly reported in
``Workshop on Quantum aspects of Black Holes'' held at Ohio State 
University, September 17-19, 2004.


\begin{thebibliography}{99}

\bibitem{ceqone}
For reviews and references to the original literature
see e.g.I.~R.~Klebanov,
``String theory in two-dimensions,''
arXiv:hep-th/9108019;
S.~R.~Das,
``The one-dimensional matrix model and string theory,''
arXiv:hep-th/9211085;
A.~Kev,
``Development in 2-d string theory,''
arXiv:hep-th/9309115;
P.~H.~Ginsparg and G.~W.~Moore,
``Lectures on 2-D gravity and 2-D string theory,''
arXiv:hep-th/9304011.


\bibitem{McGreevy:2003kb}
J.~McGreevy and H.~Verlinde,
``Strings from tachyons: The c = 1 matrix reloaded,''
JHEP {\bf 0312}, 054 (2003)
[arXiv:hep-th/0304224].

\bibitem{Klebanov:2003km}
I.~R.~Klebanov, J.~Maldacena and N.~Seiberg,
``D-brane decay in two-dimensional string theory,''
JHEP {\bf 0307}, 045 (2003)
[arXiv:hep-th/0305159].


\bibitem{Das:1990ka}
S.~R.~Das and A.~Jevicki,
``String Field Theory And Physical Interpretation Of D = 1 Strings,''
Mod.\ Phys.\ Lett.\ A {\bf 5}, 1639 (1990).

\bibitem{Minic:1991rk}
D.~Minic, J.~Polchinski and Z.~Yang,
``Translation invariant backgrounds in (1+1)-dimensional string theory,''
Nucl.\ Phys.\ B {\bf 369}, 324 (1992).


\bibitem{Moore:1992gb}
G.~W.~Moore and R.~Plesser,
``Classical scattering in (1+1)-dimensional string theory,''
Phys.\ Rev.\ D {\bf 46}, 1730 (1992)

[arXiv:hep-th/9203060].



\bibitem{Alexandrov:2002fh}
S.~Y.~Alexandrov, V.~A.~Kazakov and I.~K.~Kostov,
``Time-dependent backgrounds of 2D string theory,''
Nucl.\ Phys.\ B {\bf 640}, 119 (2002)
[arXiv:hep-th/0205079].


\bibitem{Karczmarek:2003pv}
J.~L.~Karczmarek and A.~Strominger,
``Matrix cosmology,''
JHEP {\bf 0404}, 055 (2004)
[arXiv:hep-th/0309138]


\bibitem{Karczmarek:2004ph}
J.~L.~Karczmarek and A.~Strominger,
``Closed string tachyon condensation at c = 1,''
JHEP {\bf 0405}, 062 (2004)
[arXiv:hep-th/0403169].

\bibitem{Das:2004hw}
S.~R.~Das, J.~L.~Davis, F.~Larsen and P.~Mukhopadhyay,
``Particle production in matrix cosmology,''
arXiv:hep-th/0403275.


\bibitem{Mukhopadhyay:2004ff}
P.~Mukhopadhyay,
``On the problem of particle production in c = 1 matrix model,''
arXiv:hep-th/0406029.

\bibitem{Karczmarek:2004yc}
J.~L.~Karczmarek, A.~Maloney and A.~Strominger,
``Hartle-Hawking vacuum for c = 1 tachyon condensation,''
arXiv:hep-th/0405092.

\bibitem{Karczmarek:2004bw}
J.~L.~Karczmarek, J.~Maldacena and A.~Strominger,
``Black hole non-formation in the matrix model,''
arXiv:hep-th/0411174.

\bibitem{Natsuume:1994sp}
M.~Natsuume and J.~Polchinski,
``Gravitational scattering in the c = 1 matrix model,''
Nucl.\ Phys.\ B {\bf 424}, 137 (1994)
[arXiv:hep-th/9402156].

\bibitem{Polchinski:1994mb}
J.~Polchinski,
``What is string theory?,''
arXiv:hep-th/9411028.



\bibitem{Dhar:1995hm}
A.~Dhar, G.~Mandal and S.~R.~Wadia,
``String beta function equations from c = 1 matrix model,''
Nucl.\ Phys.\ B {\bf 451}, 507 (1995)
[arXiv:hep-th/9503172].





\bibitem{Fateev:2000ik}
V.~Fateev, A.~B.~Zamolodchikov and A.~B.~Zamolodchikov,
``Boundary Liouville field theory. I: Boundary state and boundary  two-point
function,''
arXiv:hep-th/0001012.

\bibitem{Teschner:1995yf}
J.~Teschner,
``On the Liouville three point function,''
Phys.\ Lett.\ B {\bf 363}, 65 (1995)
[arXiv:hep-th/9507109].

\bibitem{Dhar:1992cs}
A.~Dhar, G.~Mandal and S.~R.~Wadia,
 ``A Time dependent classical solution of c = 1 string field theory and
nonperturbative effects,''
Int.\ J.\ Mod.\ Phys.\ A {\bf 8}, 3811 (1993)
[arXiv:hep-th/9212027].
 
\bibitem{Das:1995gd}
S.~R.~Das and S.~D.~Mathur,
 ``Folds, bosonization and nontriviality of the classical limit of 2-D string
theory,''
Phys.\ Lett.\ B {\bf 365}, 79 (1996)
[arXiv:hep-th/9507141].

\bibitem{Das:2004rx}
S.~R.~Das,
``D branes in 2d string theory and classical limits'',
[arXiv:hep-th/0401067].


\bibitem{Alexandrov:2003uh}
S.~Alexandrov,
``Backgrounds of 2D string theory from matrix model,''
arXiv:hep-th/0303190.

\bibitem{Ernebjerg:2004ut}
M.~Ernebjerg, J.~L.~Karczmarek and J.~M.~Lapan,
``Collective field description of matrix cosmologies,''
arXiv:hep-th/0405187.

\bibitem{Douglas:2003up} M.~R.~Douglas, I.~R.~Klebanov, D.~Kutasov,
J.~Maldacena, E.~Martinec and N.~Seiberg,
``A new hat for the c = 1 matrix model,''
arXiv:hep-th/0307195.

\bibitem{Takayanagi:2003sm}
T.~Takayanagi and N.~Toumbas,
``A matrix model dual of type 0B string theory in two dimensions,''
JHEP {\bf 0307}, 064 (2003)
[arXiv:hep-th/0307083].



\bibitem{Takayanagi:2004ge}
T.~Takayanagi,
``Comments on 2D type IIA string and matrix model,''
arXiv:hep-th/0408086.


\bibitem{Martinec:2003ka}
E.~J.~Martinec,
``The annular report on non-critical string theory,''
arXiv:hep-th/0305148.

\bibitem{Seiberg:2003nm}
N.~Seiberg and D.~Shih,
``Branes, rings and matrix models in minimal (super)string theory,''
JHEP {\bf 0402}, 021 (2004)
[arXiv:hep-th/0312170].

\bibitem{Fukuda:2002bv}
T.~Fukuda and K.~Hosomichi,
``Super Liouville theory with boundary,''
Nucl.\ Phys.\ B {\bf 635}, 215 (2002)
[arXiv:hep-th/0202032];
C.~Ahn, C.~Rim and M.~Stanishkov,
``Exact one-point function of N = 1 super-Liouville theory with boundary,''
Nucl.\ Phys.\ B {\bf 636}, 497 (2002)
[arXiv:hep-th/0202043].

\end{thebibliography}
\end{document}